\documentstyle[prb,multicol,aps]{revtex}

\def \bi{\bibitem}

\def\d{{\rm d}}
 \def\(({\left(}
 \def\)){\right)}
\def\bi{\bibitem}

\def \ov{\over}

\def \d{{\rm d}}

\def \nn{\nonumber}
\def \beqna{\begin{eqnarray}}
\def \eeqna{\end{eqnarray}}
\def \beq{\begin{equation}}
\def \eeq{\end{equation}}

\def \ov{\over}

\def \ol{\overline}

\def \ri{\right}

\def \l{\left}

\def \ab2{\alpha\beta^2}

 \newcommand \be {\begin{equation}}
\newcommand \bea {\begin{eqnarray} \nonumber }
\newcommand \ee {\end{equation}}
\newcommand \eea {\end{eqnarray}}
 
\newcommand \s {\sigma}

\input{epsf}
\begin{document}
\draft
\preprint{MA/UC3M/11/95}
\title{Fragile-glass behavior of a short range $p$-spin model}

\author{Diego Alvarez(*), Silvio Franz(**) and Felix Ritort(*)}
\address{(*) Departamento de Matematicas,\\
Universidad Carlos III, Butarque 15\\
Legan\'es 28911, Madrid (Spain)\\
e-mail: {\it diego@dulcinea.uc3m.es}\\
e-mail: {\it ritort@dulcinea.uc3m.es}\\
(**) International Center for Theoretical Physics\\
Strada Costiera 10\\
P.O. Box 563\\
34100 Trieste (Italy)\\
e-mail: {\it franz@ictp.trieste.it }}

\date{March 1996}
\maketitle

\begin{abstract}

\end{abstract} 
In this paper we propose a short range generalization of the $p$-spin
interaction spin-glass model. The model is well suited to test the
idea that an entropy collapse is at the bottom-line of the dynamical
singularity encountered in structural glasses. The model is studied in
three dimensions through Monte Carlo simulations, which put in evidence
fragile glass behavior with stretched exponential relaxation and
super-Arrhenius behavior of the relaxation time.  Our data are in
favor of a Vogel-Fulcher behavior of the relaxation time, related to
an entropy collapse at the Kauzmann temperature. We however encounter
difficulties analogous to those found in experimental systems when
extrapolating thermodynamical data at low temperatures. 
We study the spin glass susceptibility  investigating the behavior 
of the correlation length in the system. We find that the the 
increase of the relaxation time is not accompanied by any growth of 
the correlation length. We discuss the
scaling properties of off-equilibrium dynamics in the glassy regime,
finding qualitative agreement with the mean-field theory.


\vfill
\begin{multicols}{2}
\narrowtext
\section{Introduction}

The glassy state is very common in nature \cite{GO}. 
When it is reached from the liquid phase 
lowering the temperature, one finds 
a  dramatic increase 
of the relaxation time, and  
off-equilibrium phenomena 
can not be avoided 
on experimental time scales. This leads to non analytic 
behavior of the thermodynamic quantities, with a ``transition temperature''
that depends on the cooling rate. 
Despite its ubiquity,
the basic mechanisms underlying the common features, as well as 
the peculiarities  of the 
glassy behavior in different systems
are yet to be clarified. 

One of the most suggestive ideas in the glass theory, proposed long time 
ago by
Gibbs and Di Marzio \cite{gima}, relates the increase of the relaxation
time, and the observed finite time singularities to the existence of a
thermodynamic transition at the Kauzmann temperature where the
configurational entropy collapses to zero \cite{KAU}. Soon after, in a
refinement of the argument, Adams and Gibbs argued in favor
of a Vogel-Fulcher singularity in the relaxation time.

Disordered systems have been proposed as paradigmatic models in which
glassy phenomena can be studied in a nutshell, and theoretical ideas
tested on microscopic models \cite{kirtir}.  This is due to the fact
that in disordered systems, the glassy state already appears in
mean-field theory. The natural separation of the variables among
``quenched'' and ``annealed'' allow for the successful use of powerful
techniques as the replica method for static \cite{MPV} and functional
methods in dynamics \cite{SOZI}.  In fact,  a satisfactory
mean-field theory of disordered systems for the static \cite{MPV} as
well as for the equilibrium \cite{SOZI} and off-equilibrium
dynamics  exists \cite{cuku,frme}. Recently,
 examples of mean-field
deterministic models with glassy behavior very similar to the one of
disordered systems  have been displayed \cite{nondissis}. This points in
the direction that common mechanisms could lead to the glassy behavior of 
disordered and non-disordered systems. 

The simplest example in which the Gibbs-Di Marzio collapse occurs is the
 Random Energy Model of Derrida \cite{rem}, and is a common feature to all
systems with a ``discontinuous glassy transition'', or 
technically ``one-step replica symmetry breaking'',  where the 
Edwards-Anderson parameter \cite{EA} undergoes a discontinuity \cite{kirtir}. 
Examples of such models are
the Potts Glass \cite{grokasa}, the $p$-spin
interaction model \cite{gard} and 
a model of  manifolds in a disordered media with 
short-range correlated disorder \cite{mepa}.  
This class of systems has been proposed
by Kirkpatrick, Thirumalai and Wolynes as simple toy models for the
structural glass transition \cite{kirtir}.  Notably, the study of the
Langevin dynamics of the spherical version of these models shows that
there the Mode Coupling Theory \cite{MCT} is exact, and displays a
dynamical singularity of kind B in the G\"otze classification
\cite{MCT}.  In fact, recent progresses in the comprehension of the
dynamics of mean-field disordered systems \cite{cuku,frme} allowed for
an extension of the Mode Coupling Theory to the broken ergodicity phase
\cite{FRHE,bouetal}.

Many studies \cite{kirtir,CHS,kpv,cuku,frapa,remi}
have pointed out the existence of a
temperature $T_D$ where,  despite the fact that 
no singularity is observed in the free energy, 
there is  a statical breaking of ergodicity 
into an exponentially large number of 
metastable states. A  thermodynamic singularity is present 
at a temperature $T_C$ smaller then $T_D$. 
As a genuine mean-field theory, the Mode Coupling Theory
neglects the activated (droplet) processes that in finite-dimensional 
systems are responsible  for 
the decay of metastable states in a finite time. 
Kirkpatrick and Wolynes have recently stressed how the inclusion 
of these processes can restore ergodicity
for $T_C<T<T_D$, and  give rise to a generalized Vogel-Fulcher singularity
at the static transition temperature $T_C$ \cite{kirtir}. The argument has  
been confirmed and refined by Parisi with a theoretical calculation
based on the potential theory in spin glasses \cite{parisi}. 

The aim of this paper is to test this idea in 
a finite-dimensional disordered model where metastable states 
should be present, but destabilized
 by activated processes. In section 2 we propose a finite dimensional 
analogous of the $p$-spin interaction model in the case $p=4$. We believe
that, as usual, the mean-field limit is recovered for high
dimensionality. 
In section 3 we study the thermodynamics in the  high-temperature regime 
through Monte Carlo simulations, that demonstrate that the model behaves as
a fragile glass. 

A second aspect of our work concerns the off-equilibrium 
dynamics deep in the glassy phase. There, the properties of 
the system depend on the thermal history, and time translation invariance
does not hold \cite{Struik}. The off-equilibrium Mode Coupling Theory predicts 
scaling relations and a definite pattern of violation of the Kubo
fluctuation-dissipation relation. 
In section 4 we study the dynamics in this regime, 
showing the consistency of the aforementioned scenario. 
Finally, the conclusions are drawn.

\section{The model} 

The $p-$spin model \cite{rem,gard} is defined by the long-range  Hamiltonian 
\be
H=-\sum_{i_1<i_2<...<i_p}^{1,N} J_{i_1,i_2,...,i_p}\sigma_{i_1}
\sigma_{i_2}...\sigma_{i_p}
\label{Hmf}
\ee
where 
the couplings $J_{i_1,i_2,...,i_p}$ are independent Gaussian variables 
with zero mean, and variance $\ol{J_{i_1,i_2,...,i_p}^2}=p!/(2N^{p-1})$. 
The spins $\sigma_{i}$, $i=1,...,N$ can be taken as  Ising variables, or
as real variables subjected to the spherical constraint $\sum_{i=1}^N
\sigma_{i}^2=N$ \cite{CHS}. 
The case $p=2$, that corresponds to the Sherrington Kirkpatrick 
model, has a glassy  transition with continuous order parameter \cite{MPV} 
and will not be considered in this paper.
For $p\ge 3$  both 
in the Ising and in the spherical case the transition 
is discontinuous, and 
 the properties of the model are the ones of interest in this
paper. 
As a finite dimensional analogous of the model (\ref{Hmf}) 
in the case $p=4$ we take 
 a 
spin  system with  interacting 
Ising spins $\sigma_i$ arranged on the sites 
of a  $D$ dimensional square lattice with periodic boundary conditions. 
The Hamiltonian is defined as
\be
H=-\sum_\Box J_\Box \prod_{i \in \Box} \sigma_i
\label{H}
\ee
where the sums runs over all the plaquettes $\Box$ of the lattice.
Each spin belongs to 
$4 \l( 
\begin{array}{l}
D \\
2
\end{array}
\ri) $ 
plaquettes. Each plaquette  $\Box$ gives a contribution 
$-J_\Box \prod_{i \in \Box} \sigma_i=J_\Box \sigma_1^\Box
\sigma_2^\Box\sigma_3^\Box\sigma_4^\Box$, where the variables 
$J_\Box$ are chosen as independent Gaussian variables with zero mean and
unit variance. 
Note that, in generic dimension, the model is {\it not} invariant under
$Z^2$ gauge transformation as it would be if the spins were located on
the links (instead that on the vertices) of the plaquettes.  Thanks to
this, there is no Elitzur theorem preventing non-zero global order
parameters \cite{IT}.  In fact there are no {\it local} symmetry
operations leaving $H$ invariant \cite{foot1}. However the Hamiltonian is invariant
under the contemporary inversion of all the spins that belong to any
hyperplane of dimension $D-1$ orthogonal to one of the Cartesian axes.
It is easy to check that these operations do not change the sign of any
of the plaquettes. The degeneracy due to this symmetry (plane inversion 
symmetry in the following) is
$2^{DL-(D-1)}$, and can be removed e.g. fixing the spins on the Cartesian
axes to arbitrary values.
This exponentially large degeneracy of the states is also present 
in the ferromagnetic version of the model ($J_\Box=1$). We think
that this could lead to a very interesting spinodal dynamics,  above the 
lower critical dimension $D=2$. We concentrate here to the 
disordered model, leaving  the study of the ferromagnetic case 
for future work. A ferromagnetic model with four spin interactions
at the vertices of plaquettes was studied in \cite{maritan}
in connection with random surfaces physics. In that case 
also pair interactions  that removed
the plane inversion symmatry where present in the Hamiltonian. 

 In the limit of infinite dimension, where the
number of plaquettes to which a spin belongs tends to infinity, one can
expect that, modulo the symmetry, the model is equivalent to the
$p-$spin model for $p=4$. Models with different $p$'s could be easily
constructed for other lattices, e.g. the case $p=3$ would correspond to
a triangular lattice \cite{BAXTER}.

It is worth at this point to present a brief qualitative review of the
results of the mean-field theory, based on the Hamiltonian (\ref{Hmf})
\cite{kirtir,kpv,frapa}.  The study of the thermodynamics of this
system leads to the following results.  At high temperatures the system is
paramagnetic and ergodic.  At a temperature $T_D$ the ergodicity breaks
down, and an exponentially large number ($\exp(N\Sigma(T))$) of pure
states (ergodic components) separated by barriers of order $N$
contribute to the partition function \cite{kirtir}. This transition
occurs without singularities in the free energy, which is equal to the
free-energy per state $F_{in}$, plus an entropic contribution
$-T\Sigma(T)$ coming from the multiplicity.  The quantity $\Sigma$,
the configurational entropy, is a decreasing function of the
temperature, and at a temperature $T_C<T_D$ vanishes. At $T_C$
there is a thermodynamic phase transition with
singularities in the free-energy.

On the other hand, the  study of the off-equilibrium 
dynamics after a 
sudden quench from a high temperature shows that the 
large-time limit of various dynamical quantities is non
analytic and the singularity in the dynamics at $T_D$ is suppressed.
The characteristic time scale $\tau$ for these processes to 
restore ergodicity 
has been recently estimated in Potts-glass models
using heuristic arguments by 
Kirkpatrick and Wolynes \cite{kirtir}, and substantiated
using a droplet argument in replica space by Parisi \cite{parisi}. 
They find a generalized Adam-Gibbs relation of the kind
\be
\tau \sim \exp\l({C\ov T\Sigma(T)^{\gamma}}\ri).
\label{drop}
\ee
where $C$ is a constant and $\gamma=D-1$. 
As the configurational entropy vanishes linearly 
near $T_C$, eq.(\ref{drop}) results in a generalized Vogel-Fulcher
law $\tau \sim \exp\l({C\ov (T-T_C)^{\gamma}}\ri)$. 
The value $\gamma=2$ in $D=3$ is at variance with the usual value
$\gamma=1$ used to fit the experimantal data.
However, 
 in the case
of the present model the value of $\gamma=D-1$ 
in (\ref{drop}) should be lowered
due to the plane inversion symmetry. We 
do not know if this would result in $\gamma=D-2$, and we leave the 
investigation of this point for 
future work. As the matter of fact,  simple (and trivial) results are
obtained for the statics in $D=2$. In that case, one can show that in the high
temperature expansion only diagrams involving a number of spins
proportional to $L$ or higher, and hence irrelevant for $L\to\infty$,
are present. Accordingly the free-energy per spin is
found $F=-T \int {\d J\ov \sqrt{2\pi}}\exp(-J^2/2)\log(2\cosh(\beta
J))$. We will see that the relaxation time follows  a simple 
Arrhenius law in this case. 
 The lowest dimension at which one can expect non trivial
thermodynamical results is $D=3$, where one can see that 
frustration is present. The study of the properties of the
three-dimensional model through Monte Carlo simulations and the
comparison with the results of the theory, will be the subjects of the
rest of this paper. Some results for the two dimensional case will
 also be  mentioned.

\section{Thermodynamics}

In order to investigate the questions posed in the previous sections, 
we have performed Monte Carlo  simulations of the model in three dimensions,
using a standard serial single spin-flip heat-bath algorithm. 
The signature of glassy behaviour is easily seen in simulations of
cooling experiments. In figure 1 we plot the energy as a function of the
temperature for different cooling rates. We clearly see a change
of behaviour corresponding to a jump of the 
specific heat around $T\simeq 0.7$, where the 
systems fails to reach equilibrium within the observation 
time. The inset shows that the ``transition
temperature'' as well as the value of the energy at which the system 
freezes are dependent on the cooling rate. Since we expect the 
equilibrium entropy to be the relevant quantity for the 
transition, we integrated the high-temperature energy data 
in $\beta$ to get the entropy, taking into account
data from $\beta=0.01$ up to $\beta=1$. 

In figure 2 we present the results of this
operation, togheter with some rational function fits of the data points. 
The functional form that we have chosen to fit that data are:
\be
s_1(\beta)={\log(2)+a\beta^2\ov 1+b \beta^2}
\;\;\;\;
s_2(\beta)={\log(2)+a\beta^2\ov 1+b_2 \beta^2+b_4\beta^4}
\label{extr}
\eeq
Even functions of $\beta$ have been chosen coherently with 
the fact that 
the high temperature expansion of $S$ only contains even powers of $\beta$.
The fits are roughly  of the same quality at high temperature and are both  
 consistent with the entropy collapse 
scenario. 
Extrapolating at low temperature, one can estimate the point 
where  the entropy vanishes to be  in 
the range $0.3\le T_c\le 0.6$.  
This range of  temperatures  should not  be considered as  more 
than a mere indication of what it could happen.
First of all, the extrapolation  at low temperatures 
by means of the functions (\ref{extr}) is highly 
arbitrary. Second, we expect only the {\it configurational} entropy,
and not the total entropy (which take an intra-state contribution), 
 to become negative at the transition. 
The impossibility to disentangle the 
two contributions  adds incertitude
to the critical temperature estimate. However, both the mean-field
theory and the off-equilibrium dynamics of section 4 suggest that 
in all the low-temperature region the Edwards-Anderson parameter 
is close to 1, correspondingly the intra-state entropy is 
rather small, and possibly much smaller than the configurational
one not too close to $T_c$. 

Coherent informations are obtained from the study of the equilibrium
dynamics at high temperatures. The analysis of the autocorrelation function 
$C(t)=<\sigma(t)\sigma(0)>$ shows how the relaxation follows a 
stretched exponential law of the type,

\be
C(t)=exp(-(t/\tau)^b)
\label{e1}
\ee

As shown in figure 3 our data are pretty well fitted by the previous
relaxation law all along the relaxation. 

From that fit we are able to
extract the temperature dependent relaxation time $\tau(T)$ and the
exponent $b(T)$. The behavior of these parameters as a function of the
temperature is depicted in figure 4.  The relaxation time is
consistent with a Vogel-Fulcher law of the kind $\tau=A
\exp(B/(T-T_0))$ with $T_0\simeq 0.58$, $A=0.5$, $B=10$ (continuous
line in figure 4).  But, analogously with what happens with
experimental data, it is also well fitted by a zero temperature
singularity of the type $\tau=A \exp(B/T^2)$ with $A\simeq 11.5$,
$B\simeq 0.69$.  A four parameter fit of the form $\tau=A
\exp(B/(T-T_0)^\gamma)$ fits the data with $\gamma=1.01$, while posing
by hand $\gamma=2$ as in (\ref{drop}) leads to $T_0=0.001$.  The
indication of a thermodynamic transition by the divergence of the
relaxation time at $T_0$ is supported by the fact that it is in the
range of temperatures where the extrapolated entropy vanishes. The
exponent $b$ is also depicted in the inset of figure 4. It appears to
be linear with the temperature and of order .5 in the low temperature
phase.  Unfortunately we have no evidence stronger than this in favour
of the finite temperature singularity. The best we can say is that our
numerical data are consistent with the glass transition scenario as
much as laboratory experiments on glasses give support to this
singularity (with the difference that laboratory experiments can
explore a larger window of time than in the numerical experiments).
However the picture we get is coherent with the theoretical relation
between the relaxation time and the configurational entropy. We have
related the two quantities as suggested by eq.(\ref{drop}). In figure
5 we plot the logarithm of the relaxation time versus the inverse
entropy, and we see that the data fall quite well on a straight line
indicating the validity of (\ref{drop}) with $\gamma=1$. This linear
relation is again consitent with a small value of the intra-state
entropy, that could be estimated of the order of $S_\infty=0.04$,
value obtained from the extrapolation of the fit in fig. 5 to
$\tau\to\infty$.

It is worth at this point to mention the results of an analogous
analysis for the two dimensional system, where the thermodynamics is
trivial.  Simulations performed in that case indicate that also in $D=2$
there exists a low-temperature regime where the equilibrium correlation
function behaves as a stretched exponential (we observe that stretched
exponential behavior is also observed in the one-dimensional Ising model
at low temperatures in the intermediate time regime
\cite{BrPr}). However, the relaxation time grows much slower at low
temperature than in the three dimensional case. In fact, very good fits
are obtained by simple Arrhenius forms $\tau\sim \exp(B/T)$, i.e.  the
minimal increase to be expected in any system with discrete spin
variables.

We note that the two-step relaxation  characteristic of many structural
glasses is not seen here.


To clarify further the picture about the thermodynamics of the 
three dimensional model, 
we have investigated the possibility  of a growing static correlation 
length. This is done in disordered spin system through the analysis
 of the correlation function  \cite{ogielski}

\be
G(x)=\overline{\langle\sigma_0\tau_0\sigma_x\tau_x\rangle}\sim 
\exp(-\frac{x}{\xi})
\label{e2}
\ee

among two different replicas 
 $\sigma_i$ and $\tau_i$ with  same 
 $J_\Box$'s, and 
evolving with independent thermal noise. 
To avoid problems with the symmetry of the system 
 we have fixed all the spins $\sigma_i$ and
$\tau_i$ belonging to the Cartesian axes. In the thermodynamic limit
this should be an effective way to eliminate the degeneracy. 

In our simulations we have not found any evidence
of a growing  correlation length, which appears to be independent of
temperature and smaller than 1.
We conclude that the divergence of the relaxation time is not accompanied by
a divergent correlation length. 

An alternative way to explore the existence of a divergent correlation
length is to  compute directly the integral of the correlation function
$G(x)$ over the whole space. This quantity yields the spin-glass
susceptibility \cite{sgsusc} defined by,

\be
\chi_{SG}=N (\overline{\langle q^2\rangle}-(\overline{\langle q\rangle})^2) 
\label{e3}
\ee

where $N=L^3$ is the size of the system. For vanishing correlation
length this quantity is equal to $1$ and tends to increase whenever
there is spatial ordering and the correlation length increases
\cite{sgsusc}.  We have simulated several system sizes $L=4,5,6,7$
from $T=3.0$ down to $T=0$. Data is shown in figure 6. As one can
expect, there are serious thermalization problems especially at low
temperatures.  Nevertheless, it clearly emerges form these results
that, in the region of temperatures where we are in equilibrium (for
instance, above $T=1.2$ where the equilibrium relaxation time is
smaller than the thermalization time we made the system evolve before
measuring the observables) there is no evidence of a divergence of the
spin-glass susceptibility with the size of the
system\cite{foot2}. This is at variance with the results obtained in
case of short-range Edwards-Anderson spin glasses, where the growing
of the correlation length has been generally observed \cite{ogielski}.

\section{Off-equilibrium dynamics}

In the previous section we have seen that the relaxation time
increases very fast as the temperature is lowered, and it is equally well
fitted by a finite-time Vogel-Fulcher law and by the law 
$\tau =A\exp(B/T^2)$. Even in this second hypothesis,  in the whole 
low temperature range $T<1$, the time scales we
 can computationally  reach for reasonably large systems are  by 
far shorter than the equilibration time. History dependent 
effects and aging are then to be expected \cite{Struik}. 

Recent results in the study of the dynamics following a quench in the
low temperature phase in spin glasses, give prediction about the
scaling of the correlation function and the linear response function
for large times in the off-equilibrium regime, and on the dependence
of these quantities on the time $t_w$ that the system has spent at low
temperature \cite{cuku}.  These functions are defined respectively as
\beqna C(t,t_w)={1\ov N}\sum_i\langle \s_i(t+t_w)\s_i(t_w) \rangle
\nn\\ R(t,t_w)={1\ov N}\sum_i {\delta \langle \s_i(t+t_w) \rangle\ov
\delta h_i(t_w)} \eeqna 
where $h_i$ is a local magnetic field applied
to the system.

For a complete exposition of the off-equilibrium theory of the glassy
dynamics we refer the reader to \cite{cuku,frme}. Here we limit to
reassume briefly some features relevant to the present discussion.  For
large $t_w$ the following scenario is found: there is a first regime,
for small $t$ ($t<<\tau(t_w)$ see below) where the dynamics has features
similar to those of an equilibrium system. The correlation function is
independent of $t_w$ in this regime, and the response function is
related to the correlation function by the fluctuation-dissipation
theorem. In this regime the correlation function monotonically decreases
from 1 to a value $q_{EA}$, which defines an off-equilibrium parameter
analogous of the Edwards-Anderson parameter.  In addition to this
equilibrium-like regime, in the class of models of interest for this
paper, there is a regime in which the correlation decays from $q_{EA}$
to zero, and the correlation has the scaling form \be
C(t,t_w)=C(t/\tau(t_w)).
\label{scaling}
\ee
The ``effective relaxation time'' $\tau(t_w)$ 
is an increasing (and diverging) function of $t_w$, that the theory
is not able to predict, and seems to be rather system dependent.
In some cases it is found $\tau(t_w)=t_w$ \cite{odagaki,BM,BGI},
 we will see that this is 
not the case for the present model.  

As far as the behaviour of the response function is concerned,
it is found that the function 
\be
x(t,t_w)={T R(t,t_w)\ov {\partial C(t,t_w)\ov \partial t_w}-
{\partial C(t,t_w)\ov \partial t}},
\ee
sometimes called fluctuation-dissipation ratio,
depends on its time arguments in a quite special way:
through the dependence on $t$ and $t_w$ of the correlation function 
itself,
\be
x(t,t_w)=x(C(t,t_w)). 
\label{qfdt}
\ee
The value $x=1$ corresponds to the Fluctuation-Dissipation 
Theorem relation and it is valid for $q_{EA}<C(t,t_w)<1$.
A non zero value of $x(C)$ is found when aging effects are present 
in the response function, and 
 testifies
the memory of the system about its history.
 In  the mean-field  $p$-spin model
it is found that $x(C)$ is equal to a constant $x$ smaller than 
 1 in the whole interval $0<C<q_{EA}$. For a discussion 
of the behaviour of this quantity in the Edwards-Anderson model, 
as well as for a qualitative discussion of its behaviour in 
a different glassy scenario see \cite{frrie}. 

The response function is measured from 
 simulations of ``zero-field cooled experiments'' \cite{zfc}.
Starting at time zero from a random configuration, we  let evolve the system
at  constant temperature in zero field for a time $t_w$. 
 At  $t_w$
we switch on a small magnetic field $h$, and measure the relaxation 
of the magnetization as a function of time. In the linear response regime, 
 the magnetization $m(t,t_w)$ is given by 
\be
m(t,t_w)=h \int_{t_w}^{t+t_w} \d s \ R(t,s),
\ee
a relation that, assuming the validity of (\ref{qfdt}),
takes the form 
\be 
m(t,t_w)={h\ov T} \int_{C(t,t_w)}^1 \ \d q \ x(q) = {h\ov T}\chi(C(t,t_w)).
\ee

The mentioned behaviour of $x(C)$ reflects in 
\be 
\chi(C)=
\left\{
\begin{array}{ll}
1-C & C\geq q_{EA} \\
1-q_{EA}(1-x)-C x &\;\;\; C\leq q_{EA} 
\end{array}
\right. 
\ee

It has been noted in \cite{BBM} that, while a scaling behavior of the
kind (\ref{scaling}) is common to a glassy behavior and phenomena of
domain-growth in phase separation \cite{Bray}, the function $x(C)$ seems
to be non zero only in the aging regime of the glassy systems. In order
to discriminate glassy behavior from domain-growth like mechanisms
another quantity has been recently proposed\cite{BBM}. This is defined
considering the evolution of two replicas of the system, $\{\sigma_i\}$
and $\{\tau_i\}$ which follow identical evolution up to a time $t_w$,
and independent evolutions afterwards.  One then considers the
correlation \be Q(t,t_w)=\langle \sigma_i(t+t_w)\tau_i(t+t_w)\rangle,
\ee that, by definition, is equal to one for $-t_w\le t\le 0$.  Barrat,
Burioni and Mezard have discussed in detail the meaning of this variable
proving that $g= \lim_{t_w\to\infty}\lim_{t\to\infty} Q(t,t_w) $ is
different from zero in some domain-growth models, while it tends to zero
in mean-field spin-glasses. This shows that, in the last case, the two
typical trajectories explore different regions of the phase space. In
equilibrium, it holds the relation $Q(t/2,t_w)=C(t,t_w)$ (with both
quantities independent of $t_w$).  In has been also shown that the same
relation holds in a trap model even in the aging situation.

We have investigated the behavior of the functions $C$, $R$ and $\chi$
in numerical simulations for a large size $L=30$ and temperatures
$T=0.5,0.7$. The two temperatures give qualitatively similar results. 
In figure 7 we plot our data for $C(t,t_w)$ and $Q(t,t_w)$. As anticipated,
the scaling $\tau(t_w)=t_w$ is not obeyed. A best fit of the 
form $\tau(t_w)=t_w^\alpha$ gives $\alpha=0.77$ and produce a fairly
good collapse of the data on the same master curve. We did not 
investigate a possible dependence of $\alpha$ on the temperature. 
The behavior of $Q(t,t_w)$ indicates that the parameter $g$ 
is zero in this model as predicted by the mean-field theory. 
 We see that the 
relation $Q(t/2,t_w)=C(t,t_w)$ is obeyed at short times, and 
does not work so badly even at large times. However a rescaling 
of the kind $Q(t/(1.5),t_w)=C(t,t_w)$ fits better the data.

As far as the function $x(C)$ is concerned, we plot in figure 
8 the rescaled  magnetization $\chi=(T/h) m(t,t_w)$ versus 
$C(t,t_w)$ for different $t_w$. The figure shows clearly that
for the waiting times here considered we are vary far from an asymptotic
regime where $\chi$ is independent of $t_w$. However, the slope 
of the curve for small $C$ seems not to vary too much with
$t_w$, and it is roughly equal to $x=0.4$. The region where 
$\chi(C)=1-C$ terminates roughly at a value $q_{EA}=0.97$. 
Assuming (as supported by mean-field theory) that the Edwards-Anderson
parameter defined in this way and the one defined in statics take 
equal or comparable values, we argue that the quasi-states dominating 
the thermodynamics are quite ``narrow'', and, as announced, the 
intra-state entropy is quite small. To our knowledge, the function 
$x(C)$, or equivalently $\chi(C)$ has never been measured in 
experiments. Its determination, which would involve independent 
measures of a time correlation function and its associated  response function, 
would be a good test of the spin-glass scenario in glasses. 

\section{Conclusions}

In this work we have investigated the glassy behavior of a short-range
disordered $p$-spin interaction model. We have focused to the case $p=4$
in three dimensions, reporting for comparison purpose 
some results for the two-dimensional system. 
The mean-field version of this model displays
a static transition where the configurational entropy is nearly zero and
replica symmetry breaks. We have found that the finite 
dimensional model shows interesting
features, archetypical of real laboratory glasses. Both in two and in three
dimensions the equilibrium autocorrelation function follows a stretched
 exponential form at low enough temperature. While in two dimensions 
the relaxation time increases according to the Arrhenius law, 
in dimension three our data demonstrate  a faster increase. This is compatible
both with a Vogel-Fulcher law, and with an $\exp(A/T^2)$ behavior. 
In favor of the finite-temperature transition scenario, we can only  offer
the extrapolations of the high temperature data of the 
entropy. A better support to the theory is furnished 
by  the relation $\tau\sim \exp(A/(T S(T))$ that we observe 
in the temperature window we explored. The search in several 
quantity, of a growing statical correlation length associated to 
the growing relaxation time gave negative answer. 
In all cases we found very small, and nearly temperature 
independent correlation lengths. Of course we
can not exclude that the correlation length begins to grow at temperatures
lower that the ones at which we could equilibrate the system. 

The study of the off-equilibrium dynamics of the system confirmed the
qualitative features predicted by Mean-Field Theory. The behavior of
the auto-correlation function shows aging. As times goes by the
dynamics becomes slower. The effective relaxation time, defined as the
characteristic time for the correlation function to vanish, grows as a
power of the waiting time. The analysis of the function $Q(t,t_w)$
shows that typical trajectories which coincide at time $t_w$ explore
uncorrelated regions of the space at later times.  The behavior of the
fluctuation-dissipation ratio, displays the qualitative features
expected, showing long-range memory effects in the aging
regime. However the residual dependence on $t_w$ indicates that at the
times we have reached the eventual asymptotic behavior is still very
far \cite{Bonilla}.

The evolution 
of glassy systems is often described phenomenologically as
rare jumps in a landscape of  ``traps'' (metastable states)
inside which most of the time is spent \cite{odagaki}. 
This picture matches quite well with the theoretical idea of 
the metastable states destabilized by activated processes. 
It has been pointed out in \cite{BBM} that
jumps among {\it uncorrelated} traps imply the relation 
 $Q(t,t_w)=C(2t,t_w)$.
 Our data show $Q(t,t_w)$ is indeed close to 
$C(2t,t_w)$. If this would be confirmed by  more precise and  systematic 
studies, would be an evidence in favor of the trap models.  
In this respect it would also be useful the analysis of 
the self-averaging properties of some local quantity.  

We would like to conclude mentioning some of the problems left 
open by this work.
 For example it would be interesting to study
the behavior of the dynamic Edwards-Anderson parameter as a function of 
temperature, or the large-time decay of the energy to its asymptotic value. 
We have also seen, that the equilibrium  relaxation 
in high temperature does not seem to 
 proceed in two steps ($\beta$ and $\alpha$)
as it commonly observed  in laboratory, and as it is predicted 
by Mode Coupling Theory.
As far as we know, the $\beta$-relaxation process
has never been observed in a spin system. 
We do not know if this is just due to
a difficulties to disentangle the two processes due 
 the high value of the 
Edwards-Anderson parameter in the neighborhood of  the 
transition, or to the real absence of the $\beta$ process. 
We think however  that the
understanding of this point can shed some light on the nature
of the $\beta$-relaxation process. On the theoretical side, it would be 
very interesting to understand if the observed 
value of the exponent $\gamma=1$ in three dimensions, in contrast with
the replica theory prediction for the Potts-glass $\gamma=2$ has
to be imputed to the plane inversion symmetry, or if it is observed even in
absence of this.
In that respect simulations of the Potts-glass \cite{potts} could give  
important hints.

\begin{center}
{\bf Acknowledgments}
\end{center}
We would like to thank
F. Cesi, G. Parisi,  M. M\'ezard,
 F. G. Padilla and M.A. Virasoro for 
many  discussions and suggestions. 
F.R. acknowledges Ministerio de Educacion
y Ciencia of Spain for financial support. S.F. is grateful for the 
hospitality of the  Departamento de Matematicas
of the Universidad Carlos III (Madrid) where this work was taken to
accomplishment.

\vfill\eject

\begin{center}{\bf Figure Captions}\end{center}

\begin{itemize}

\item[Fig.~1.] The energy as a function of the temperature for 
different cooling rates $r$. It is apparent the calorimetric
glass transition  around $T\approx 0.7 $. The region around the transition
is magnified in the inset. 

\item[Fig.~2.] The high-temperature entropy (data was collected for
one sample with lattice size $L=10$) versus the inverse temperature,
together with the rational fits mentioned in the test. The fitting
parameters are, respectively for the $s_1$ and $s_2$ forms: $a=-0.302$,
$b=1.866$ and $a=-0.039$, $b_2=2.060$, $b_4=1.550$.

\item[ Fig.~3.] The correlation function for various temperatures: 
$T=2.8$ ($\Box$), $T=2.4$ ($\times$), $T=2.0$ ($\bigtriangleup$), 
$T=1.6$ ($\bigcirc$), $T=1.2$ ($+$). The lines are stretched exponential
fits. 

\item[Fig.~4.] The relaxation time as a function of the temperature 
toghether with the Vogel-Fulcher fit (dotted line) and
 the $\exp(\beta^2)$ fit (slashes). The inset shows the temperature 
dependence of the exponent $b$. 

\item[Fig.~5.] The inverse of the logarithm of the relaxation time 
versus the $T S(T)$. The linear dependence is the one 
predicted by the Adam-Gibbs form. 

\item[Fig.~6.] Spin-glass susceptibility $\chi$ for different sizes
$L=4,5,6,7$ averaged over 100 samples versus the temperature.
There is no evidence of a growing correlation length.

\item[Fig.~7.] The functions $C$ and $Q$ at $T=0.5$. 
We plot on the same graph 
$C(t,t_w)$ and $Q(t/2,t_w)$. We see that the relation 
$C(t,t_w)=Q(t/2,t_w)$ is quite well obeid even for small 
values of $C$. The curves for $t_w=10,100,1000,10000$ are depicted. 

\item[Fig.~8.] The function $\chi(C)$ for $T=0.7$ and
waiting times $t_w=10,100,1000$.
The curves show a strong dependence on $t_w$ indicating 
that we are far from the eventual asymptotic regime. 

\end{itemize}

\end{multicols}
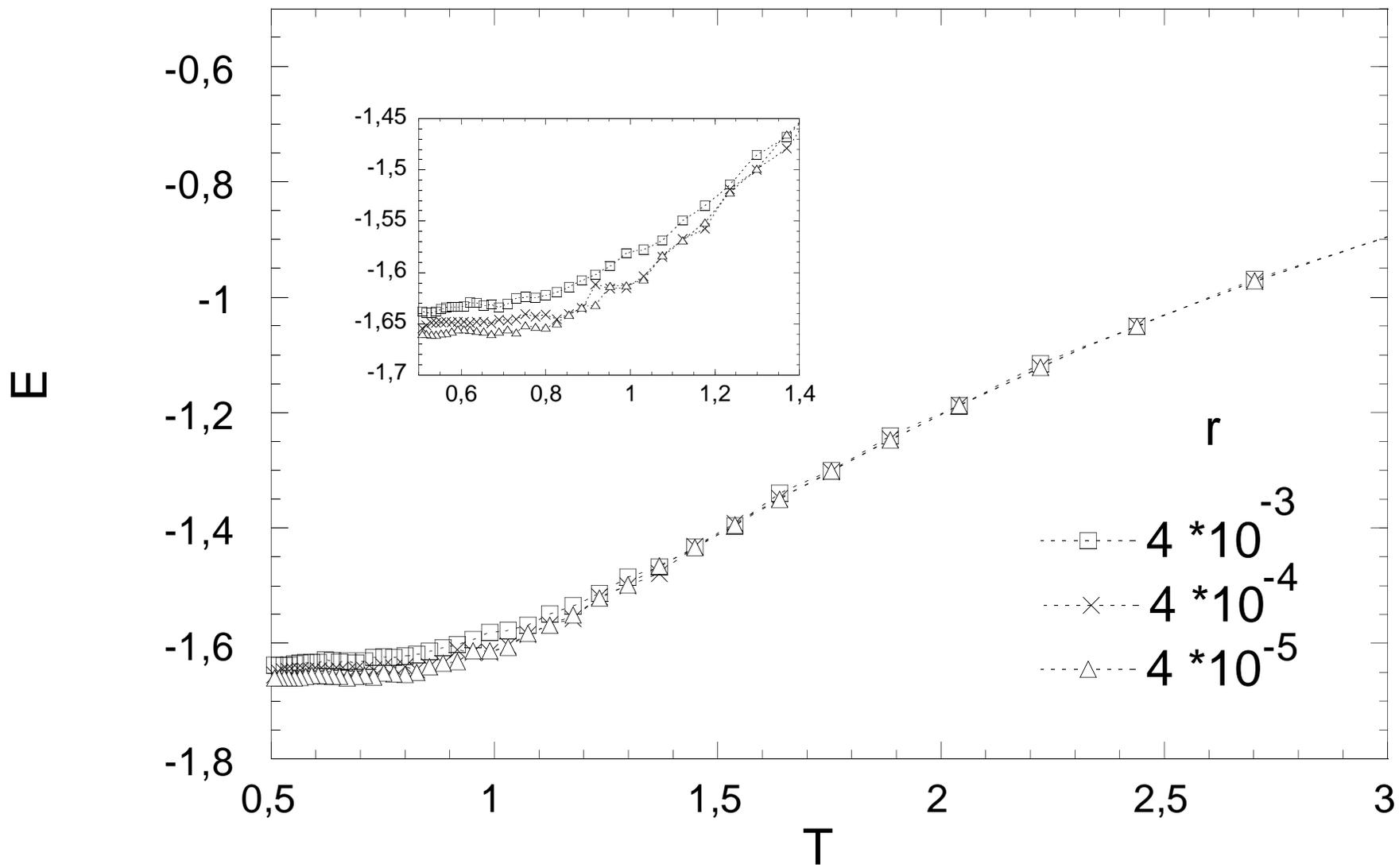
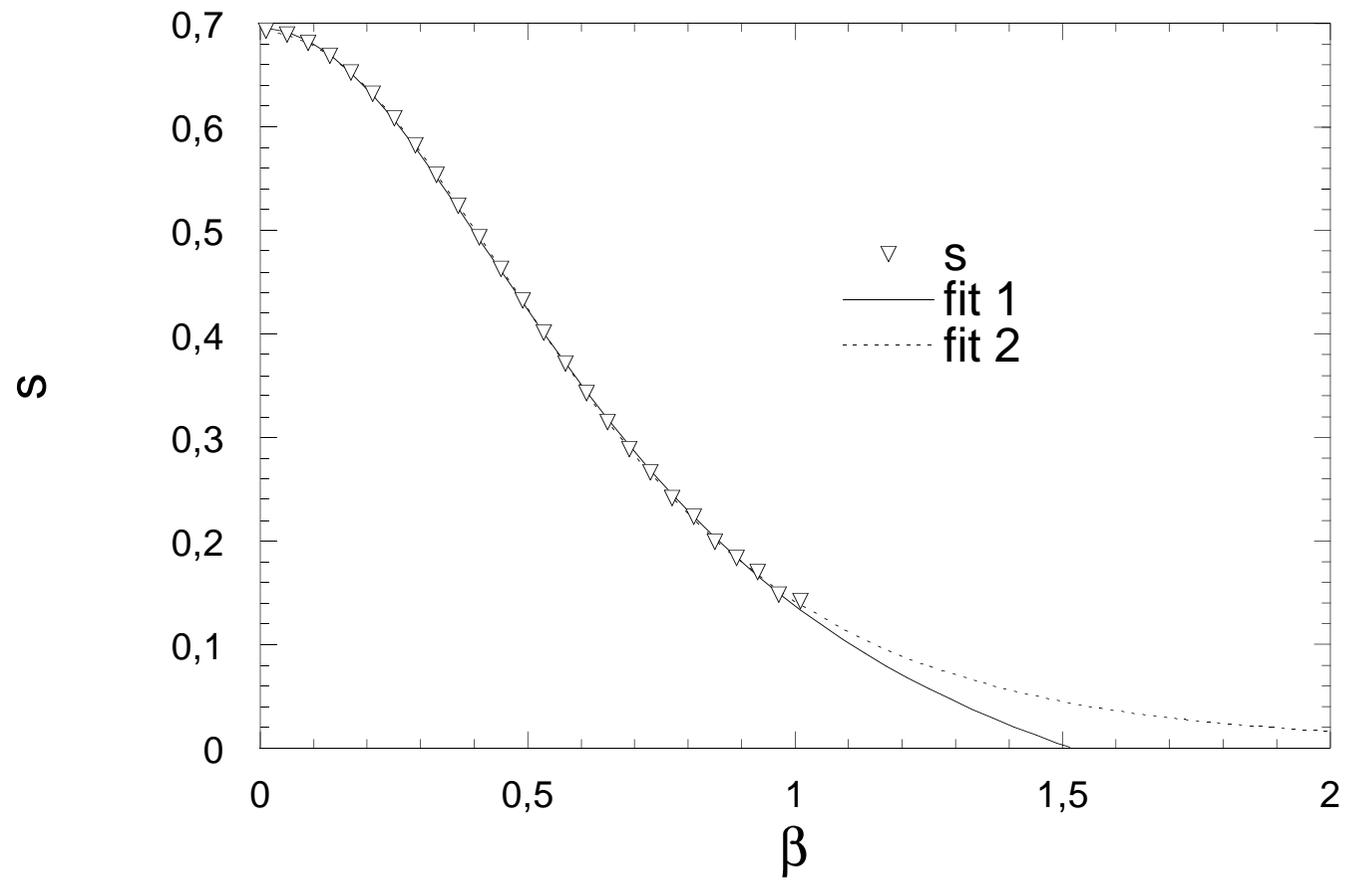
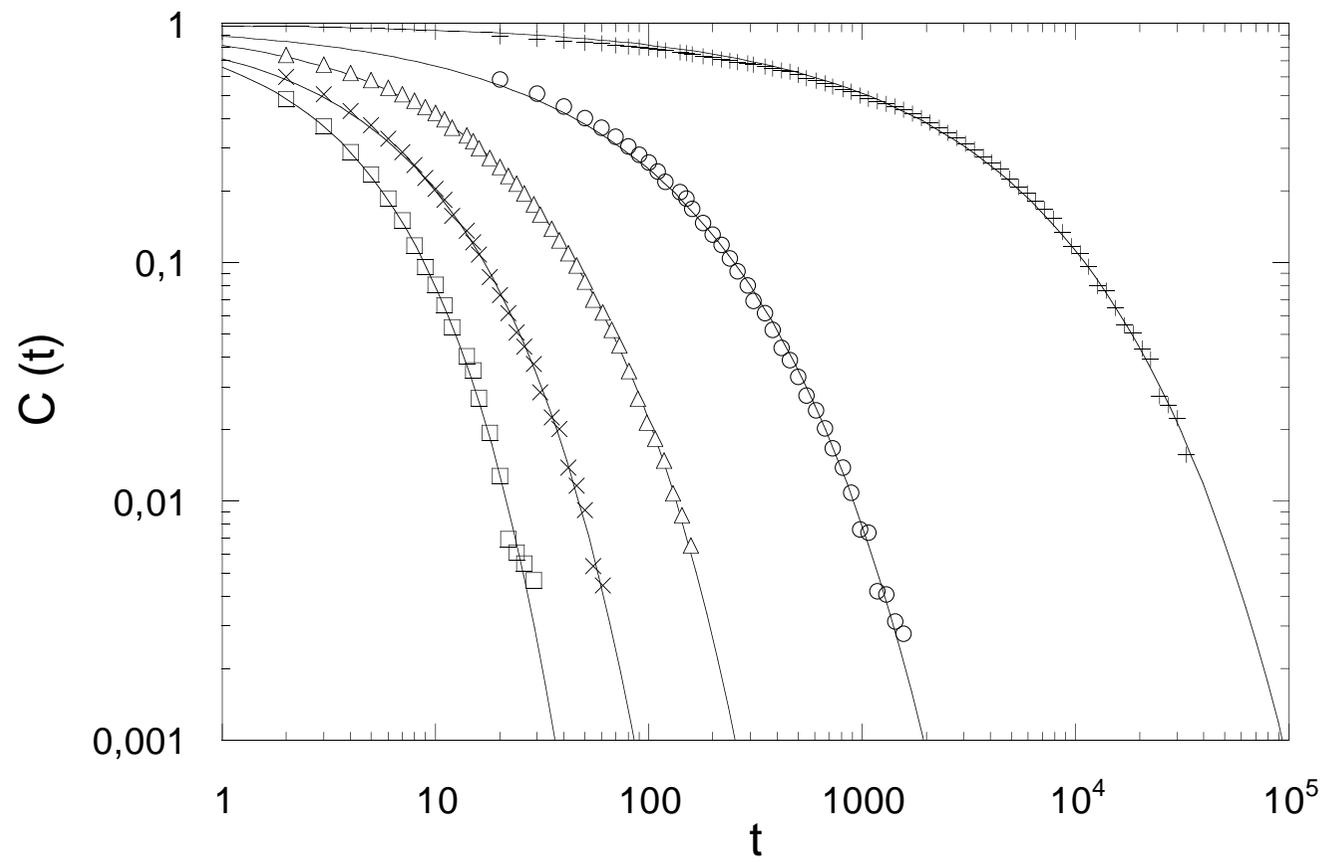
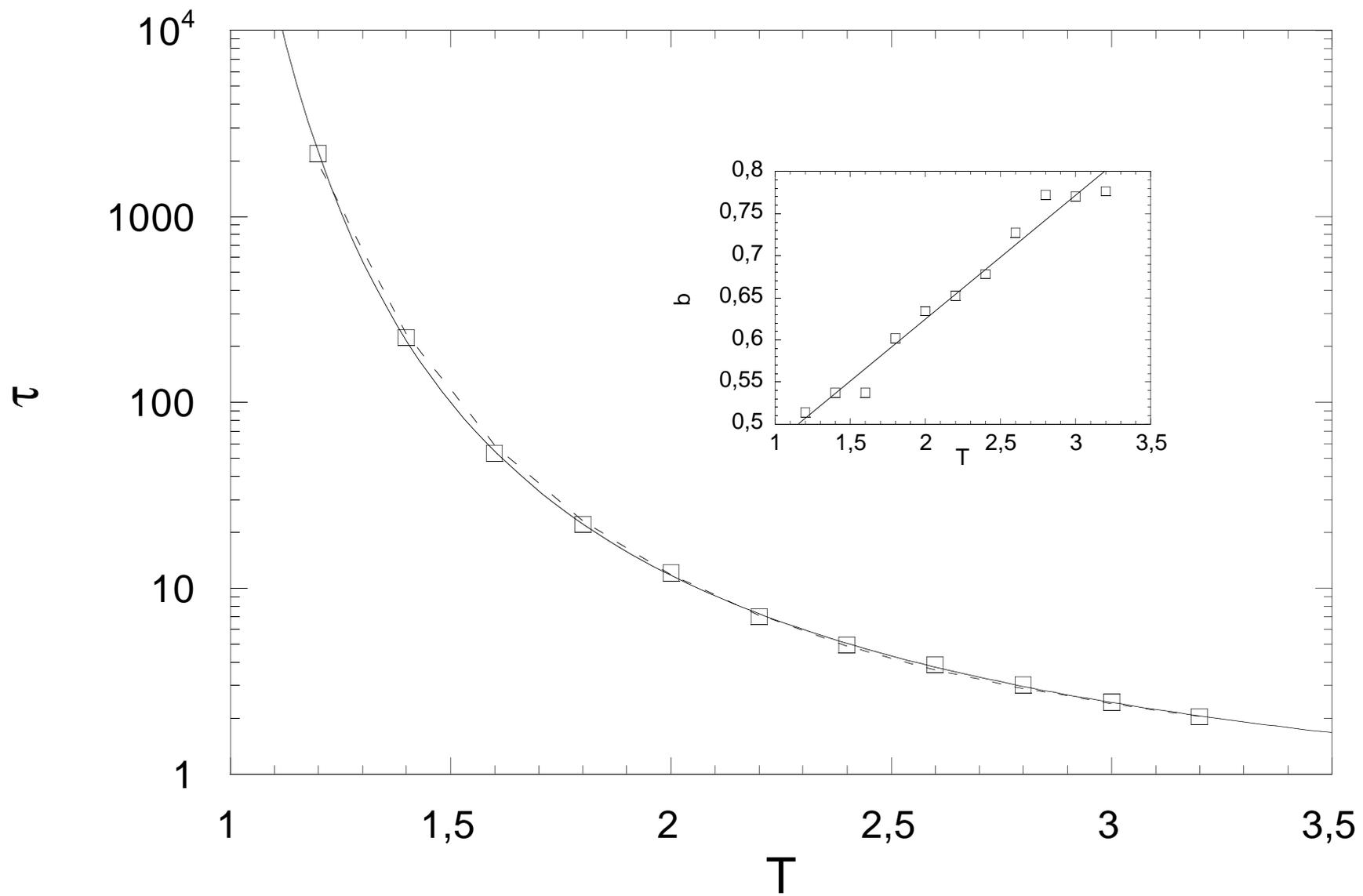
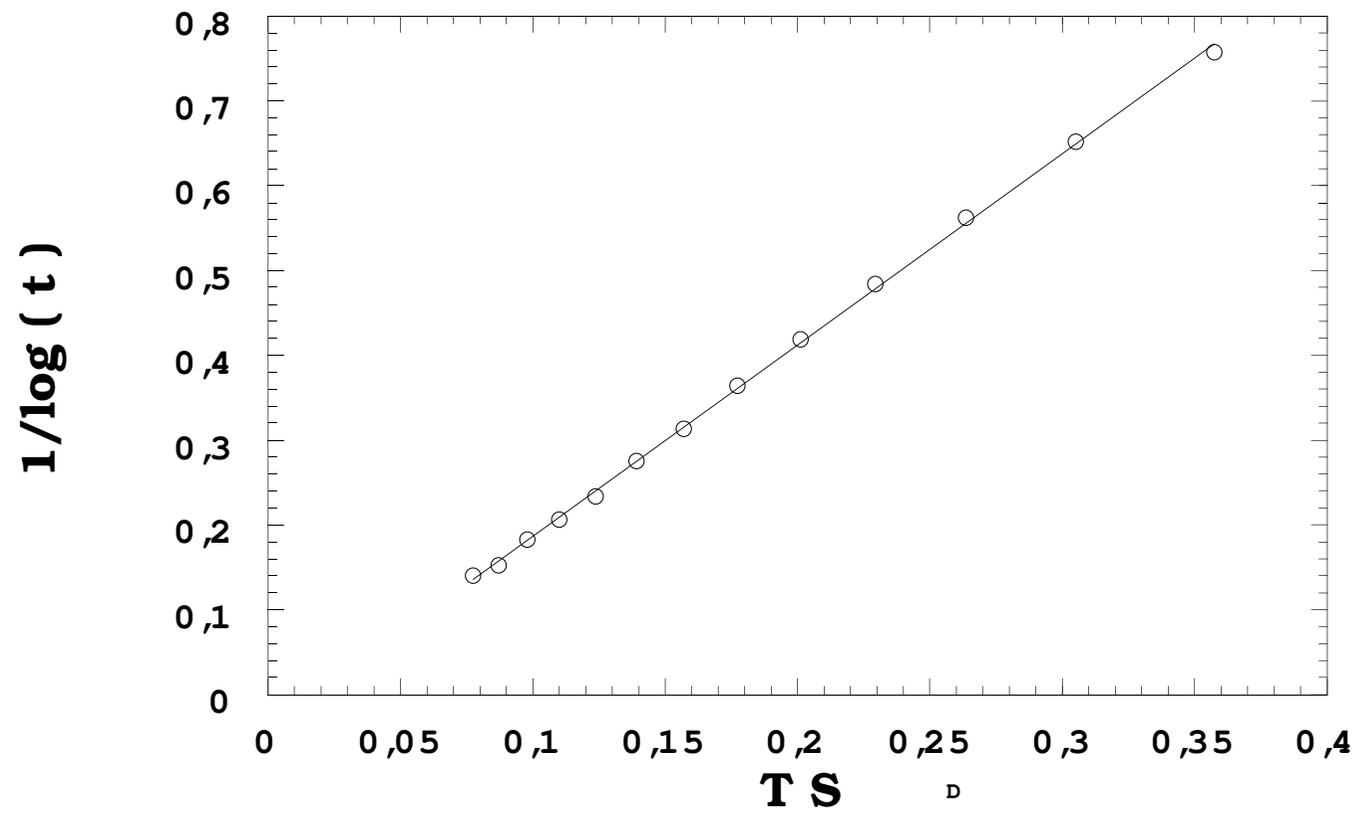
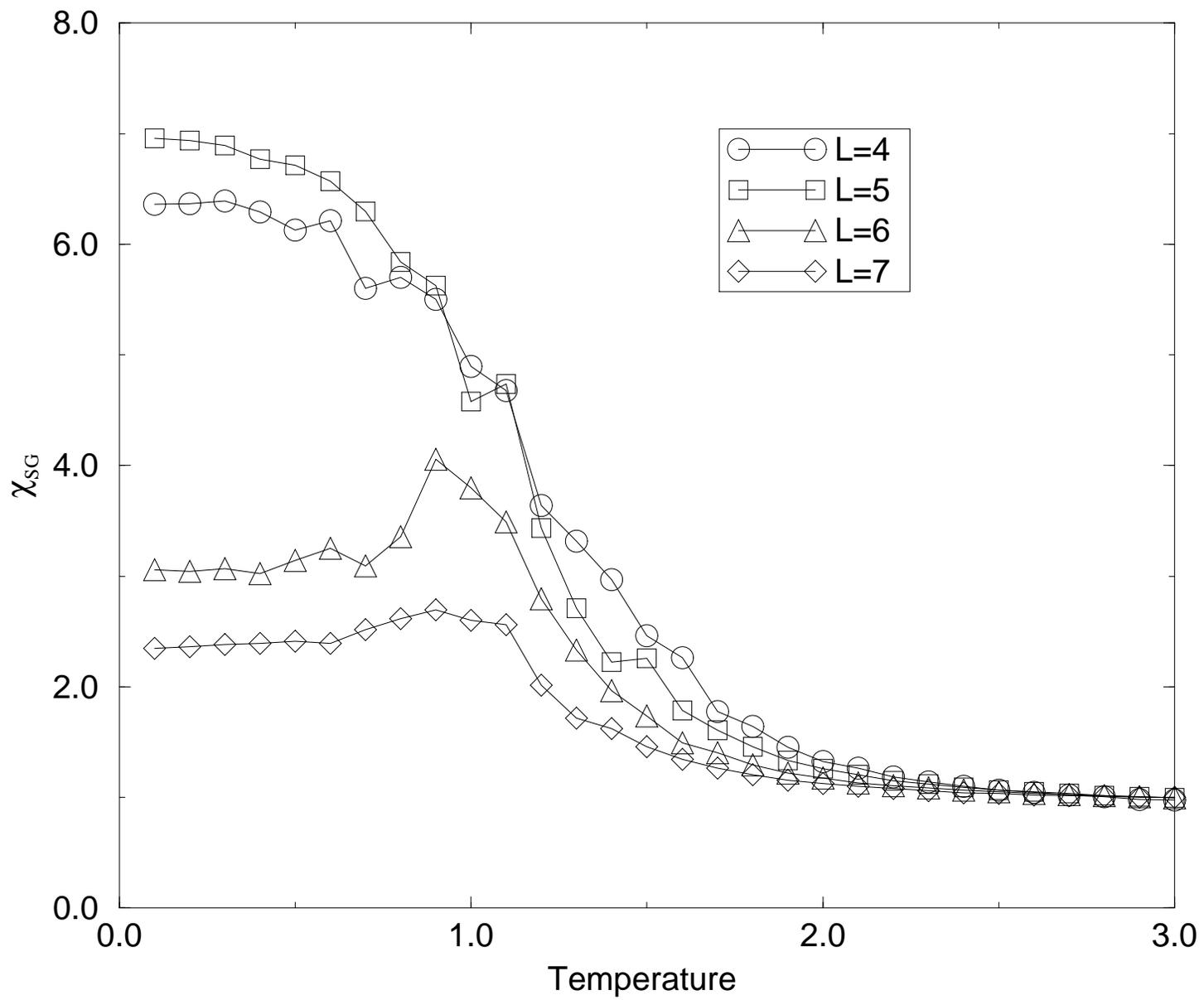
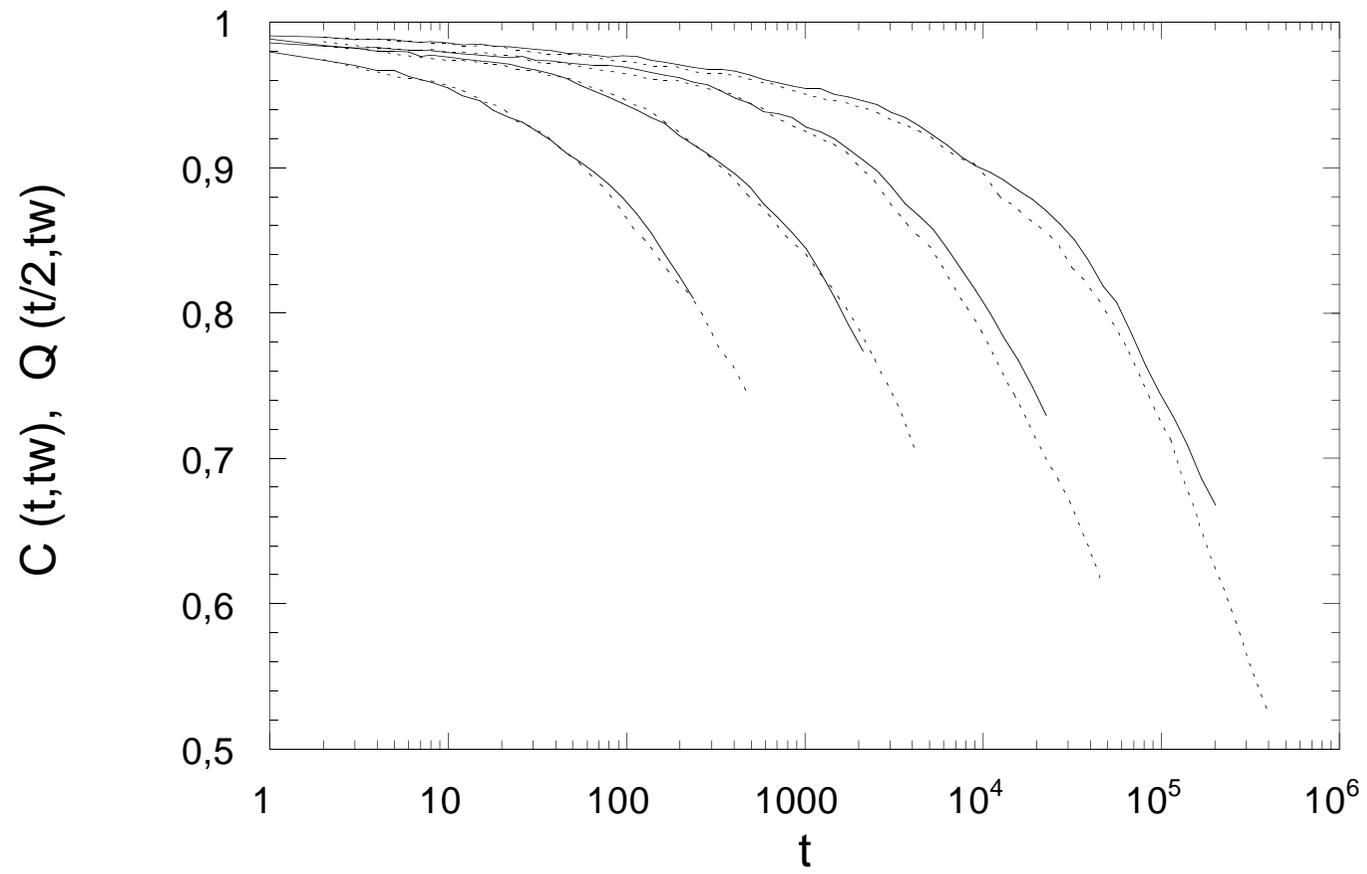
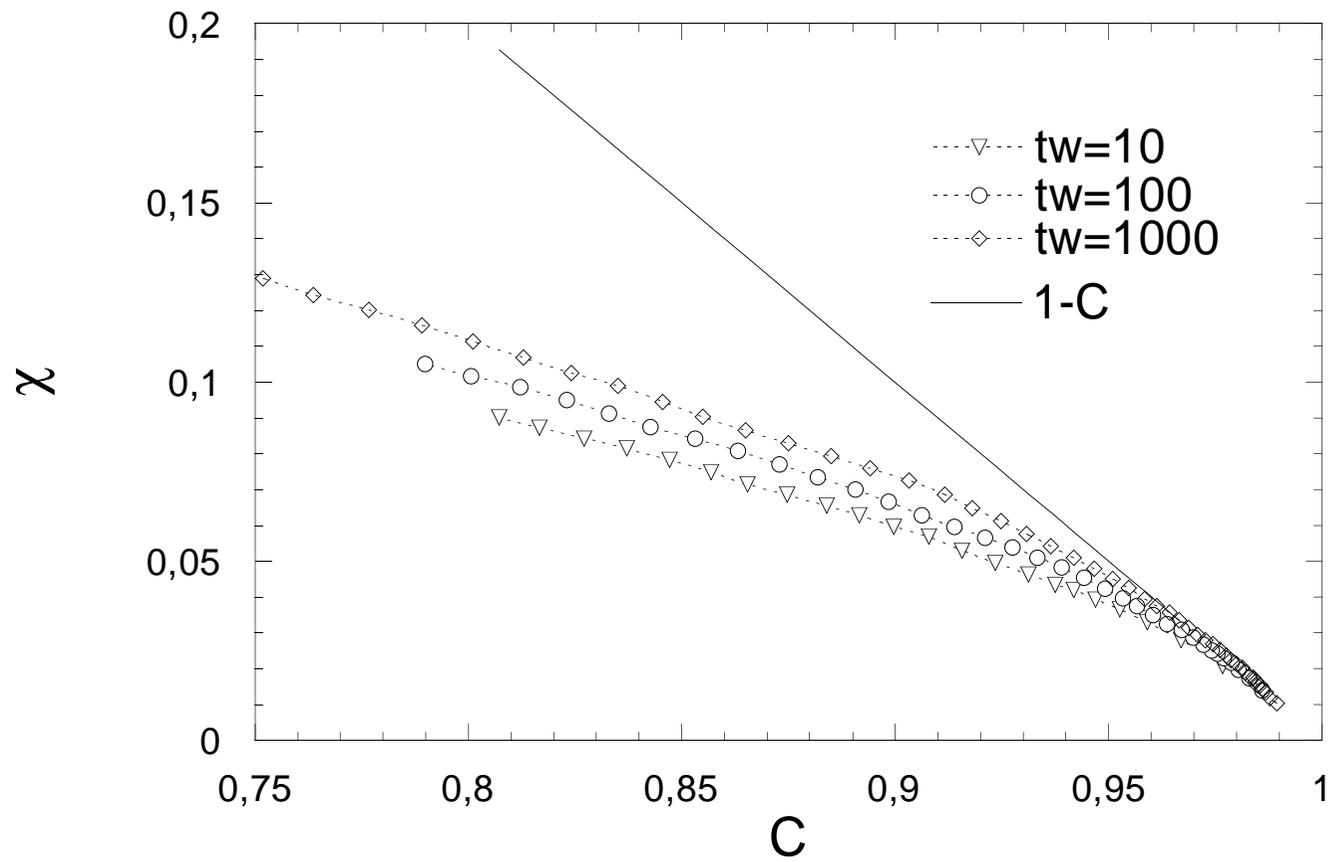
\end{document}